\documentclass[3p,times,twocolumn]{elsarticle}

\usepackage{ecrc}

\volume{00}

\firstpage{1}

\journalname{Nuclear Physics B Proceedings Supplement}

\runauth{B. Betz et al.}

\jid{nppp}

\jnltitlelogo{Nuclear Physics B Proceedings Supplement}

\usepackage{amssymb}

\usepackage[figuresright]{rotating}

\begin{document}

\begin{frontmatter}



\dochead{}

\title{Constraints on the Jet-Medium Coupling from Measurements at RHIC and LHC}


\author[label1]{Barbara Betz}
\author[label1]{Florian Senzel}
\author[label1]{Carsten Greiner}
\author[label2]{Miklos Gyulassy}

\address[label1]{Institute for Theoretical Physics, Johann Wolfgang 
Goethe-University, 60438 Frankfurt am Main, Germany}
\address[label2]{Department of Physics, Columbia University, New York, 
10027, USA}

\begin{abstract}
The measured data on the nuclear modification factor for pions and reconstructed jets 
as well as on the high-$p_T$ elliptic flow at RHIC and LHC energies are compared to
results from a linear pQCD and a highly non-linear hybrid AdS holographic model of 
jet-energy loss. We find that the high-$p_T$ ellitic flow requires to include realistic 
medium transverse flow fields and a jet-medium coupling including the effects of
the energy of the jet, the temperature of the bulk medium, and 
non-equilibrium effects close to the phase transition. We extend our jet-energy loss model that 
is coupled to state-of-the-art hydrodynamic prescriptions to backgrounds generated by the 
parton cascade BAMPS. We demonstrate that the results for the hydrodynamic and the parton-cascade backgrounds
show a remarkable similarity. Unfortunately, the results for both the pion and a parton-jet nuclear 
modification factor are insensitive to the jet-path dependence of the models 
considered. 

\end{abstract}


\end{frontmatter}

\section{Introduction}

One of the formidable tasks in heavy-ion physics is to identify a precise understanding 
of the jet-medium dynamics, the jet-medium interactions, and the jet-energy loss formalism.
Below, we study the influence of the details of the jet-medium coupling and the 
medium background on the simultaneous description of the nuclear 
modification factor ($R_{AA}$) and the high-$p_T$ elliptic flow ($v_2$) measured 
at RHIC and LHC \cite{data1,data2,data3,data4} for a radiative pQCD energy-loss ansatz \cite{us}. 
We contrast media determined via the viscous hydrodynamic approach VISH2+1 \cite{VISH2+1} 
with the parton-cascade BAMPS \cite{BAMPS} as well as a jet-medium coupling
depending on the collision energy with a jet-medium coupling influenced by 
the energy of the jet, the temperature of the medium and non-equilibrium effects 
around the phase transition. 

Besides this, we compare the jet-energy loss based on radiative pQCD \cite{us}
with the Hybrid AdS energy-loss ansatz of Ref.\ \cite{Casalderrey-Solana:2014bpa}. We
contrast the pion nuclear modification factor obtained via the radiative pQCD-energy loss 
\cite{us} and the Hybrid AdS energy-loss ansatz with a parton-jet nuclear modification 
factor that can be considered as an idealized LO Jet $R_{AA}$ at RHIC and LHC energies.

The pQCD-based energy loss model studied is parametrized as \cite{us}
\begin{eqnarray}
\hspace*{-0.5cm}
\frac{dE}{dx}=\frac{dE}{d\tau}(\vec{x}_0,\phi,\tau)= 
-\kappa\,  E^a(\tau) \, \tau^{z} \, e^{c=(2+z-a)/4} \, \zeta_q \, v_f
\;,
\label{Eq1}
\end{eqnarray}
with the jet-energy dependence $a$, the path-length dependence $z$, and the energy
dependence $c$. In the following, the jet-medium coupling $\kappa$ will 
depend either on the collision energy $\kappa=\kappa(\sqrt{s_{NN}})$ or 
the energy of the jet and the temperature of the background medium considered 
$\kappa=\kappa(E^2,T)$. The jet-energy loss fluctuations are distributed via
$f_q(\zeta_q)= \frac{(1 + q)}{(q+2)^{1+q}} (q + 2- \zeta_q)^q $,
allowing for an easy interpolation between non-fluctuating ($\zeta_{q=-1}=1$), 
uniform Dirac distributions and distributions increasingly skewed towards small 
$\zeta_{q>-1} < 1$.

The jets are spread according to a transverse initial profile specified by 
the bulk flow fields given by the VISH2+1 and BAMPS backgrounds
considered \cite{VISH2+1,BAMPS}.

\begin{figure*}[t]
\begin{center}
\includegraphics*[width=14.5cm]{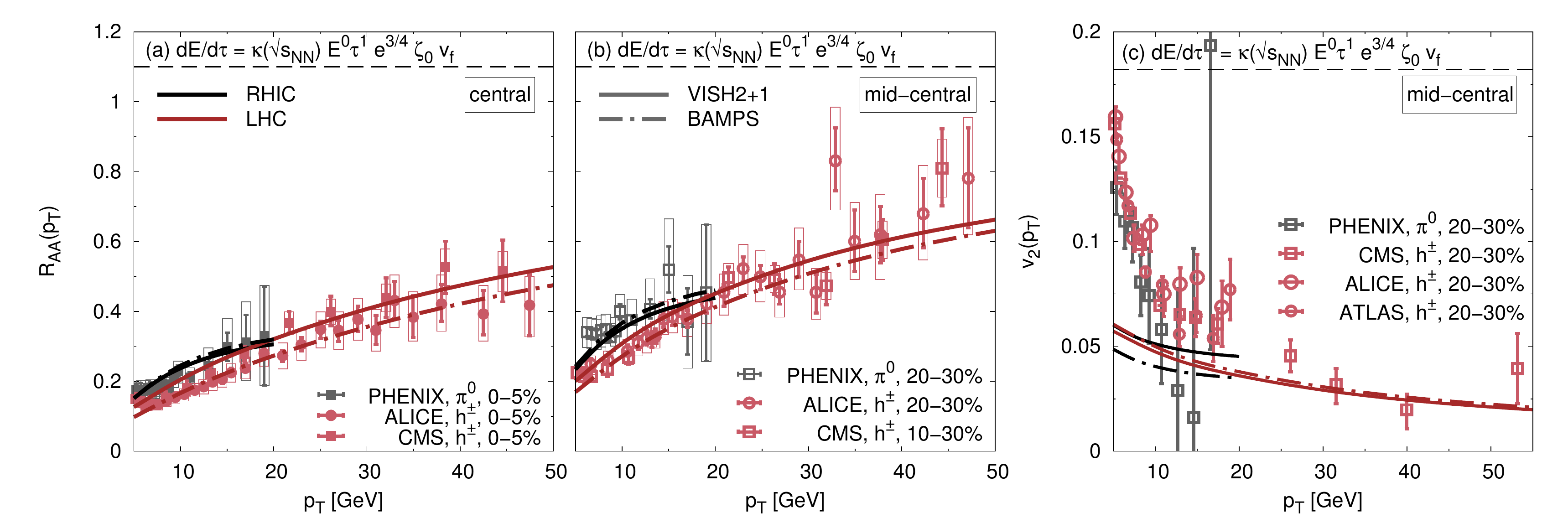}
\caption{The pion nuclear modification factor for central (left panel) and mid-central 
(middle panel) collisions at RHIC (black) and LHC (red) as well as the high-$p_T$ pion
elliptic flow for mid-central events (right panel). The measured data 
\cite{data1,data2,data3,data4} is compared to a pQCD-based energy loss 
$dE/d\tau=\kappa(\sqrt{s_{NN}}) E^0 \tau^1 e^{3/4} \zeta_{0} v_f$, including jet-energy loss fluctuations 
($\zeta_0$) and transverse flow fields ($v_f$), for a jet-medium coupling that depends 
on the collision energy ($\kappa=\kappa(\sqrt{s_{NN}})$) applying the hydrodynamic backgrounds of VISH2+1 \cite{VISH2+1} 
(solid lines) and the parton cascade BAMPS \cite{BAMPS} (dashed-dotted lines).
\vspace*{-0.4cm}}
\label{Fig01}
\end{center}
\end{figure*}

On the other hand, the jet-energy loss of the Hybrid AdS energy-loss ansatz \cite{Casalderrey-Solana:2014bpa}
is based on falling strings \cite{Chesler:2014jva} where
\begin{eqnarray}
\frac{1}{E_{\rm in}}\frac{dE}{dx} = -\frac{4}{\pi}\frac{x^2}{x^2_{\rm stop}}
\frac{1}{\sqrt{x^2_{\rm stop}-x^2}}\,.
\label{Eq2}
\end{eqnarray}
The initial jet energy is given by $E_{in}$ and the string stopping distance for quark and gluon 
jets is determined via $x^{q,g}_{\rm stop}=1/(2\kappa_{\rm sc}^{q,g})E_{\rm in}^{1/3}/T^{4/3}$
with the jet-medium coupling $\kappa_{\rm sc}^{q}=\kappa_{\rm sc}$ for quarks 
and $\kappa_{\rm sc}^{g}=\kappa_{\rm sc}(C_A/C_F)^{1/3}$
for gluons, including the respective Casimir operators $C_A$ and $C_F$. 

This energy loss ansatz has been integrated into our existing model \cite{us}. 
Please note that Ref.\ \cite{Casalderrey-Solana:2014bpa} uses natural units, 
$\hbar=c=1$. For a direct comparison, we quote our results below using a 
dimensionless coupling. 

The main differences between the two energy-loss descriptions is the square-root
dependence that leads to the formation of a Bragg peak with the explosive burst 
of energy close to the end of the jet's evolution. There have been discussions
in literature \cite{Casalderrey-Solana:2014bpa,Betz:2008ka,Ficnar} on the impact
of the Bragg peak. In line with previous findings \cite{Betz:2008ka}
we will show below that there is a difference between the 
Hybrid AdS energy-loss ansatz featuring a Bragg peak and the pQCD
model without a Bragg peak, however, this difference is only marginal.

\section{Results and Discussion}

Fig.\ \ref{Fig01} shows the pion nuclear modification factor ($R_{AA}$) for central (left panel)
and mid-central (middle panel) collisions at RHIC (black) and LHC (red)
as well as the high-$p_T$ elliptic flow ($v_2$) for mid-central events (right).
The measured data \cite{data1,data2,data3,data4} is compared to the pQCD-based
energy loss of Eq.\ (\ref{Eq1}) with $(a=0, z=1, c=3/4)$. Jet-energy loss fluctuations 
($\zeta_0$) and the transverse expansion of the background flow ($v_f$) are 
included, as well as a running jet-medium coupling that depends on the energy of the collision, $\kappa=\kappa(\sqrt{s_{NN}})$.

\begin{figure*}[t]
\begin{center}
\includegraphics*[width=14.5cm]{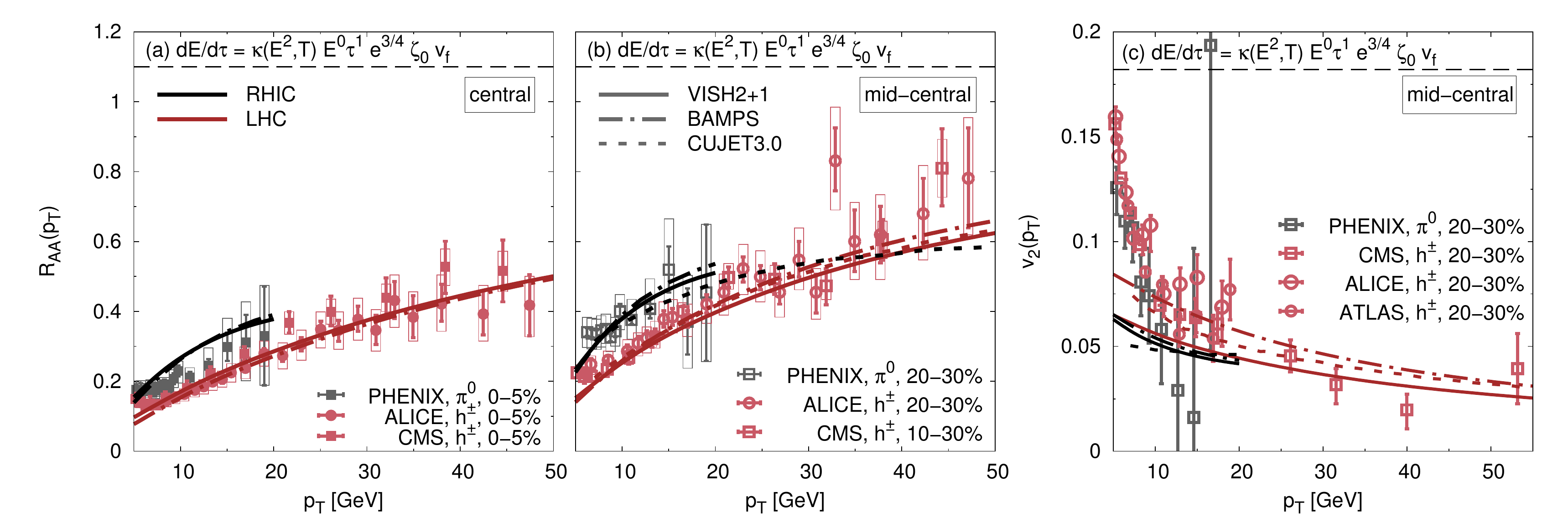}
\caption{The pion nuclear modification factor for central (left panel) and mid-central 
(middle panel) collisions at RHIC (black) and LHC (red) as well as the high-$p_T$ pion
elliptic flow for mid-central events (right panel). The measured data 
\cite{data1,data2,data3,data4} is compared to a pQCD-based energy loss 
$dE/d\tau=\kappa(E^2,T) E^0 \tau^1 e^{3/4} \zeta_{0} v_f$, including jet-energy loss fluctuations 
($\zeta_0$) and transverse flow fields ($v_f$), for a jet-medium coupling that depends 
on the energy of the jet, the temperature of the medium, and non-equilibrium
effects around the phase transition ($\kappa=\kappa(E^2,T)$) applying the hydrodynamic 
backgrounds of VISH2+1 \cite{VISH2+1} (solid lines) and the parton cascade BAMPS \cite{BAMPS}
(dashed-dotted lines). For comparison, we include the results of CUJET3.0 \cite{Xu:2014tda}
(dotted lines).
\vspace*{-0.4cm}}
\label{Fig02}
\end{center}
\end{figure*}

Fig.\ \ref{Fig01} demonstrates that there is a surprising similarity between the results
that cannot be expected a priori given the fact that the two background media are so 
different: While the hydrodynamic description of VISH2+1 \cite{VISH2+1} assumes an 
equilibrated system, the parton cascade BAMPS \cite{BAMPS} also includes non-equilibrium effects
in the bulk medium evolution.

In addition, the figure exhibits the so-called high-$p_T$ $v_2$-problem \cite{us}:
The high-$p_T$ elliptic flow below $p_T \sim 20$~GeV is about a factor of two below
the measured data \cite{data1,data2,data3,data4}. This effect has been discussed
in literature \cite{us,Xu,Molnar} and recently it has been suggested by CUJET3.0
\cite{Xu:2014tda} that a temperature and energy-dependent jet-medium coupling 
$\kappa=\kappa(E^2,T)$, which includes non-perturbative effects around the phase 
transition of $T_c \sim 160$~MeV, can overcome this problem.

\begin{figure*}[t]
\begin{center}
\includegraphics*[width=9.65cm]{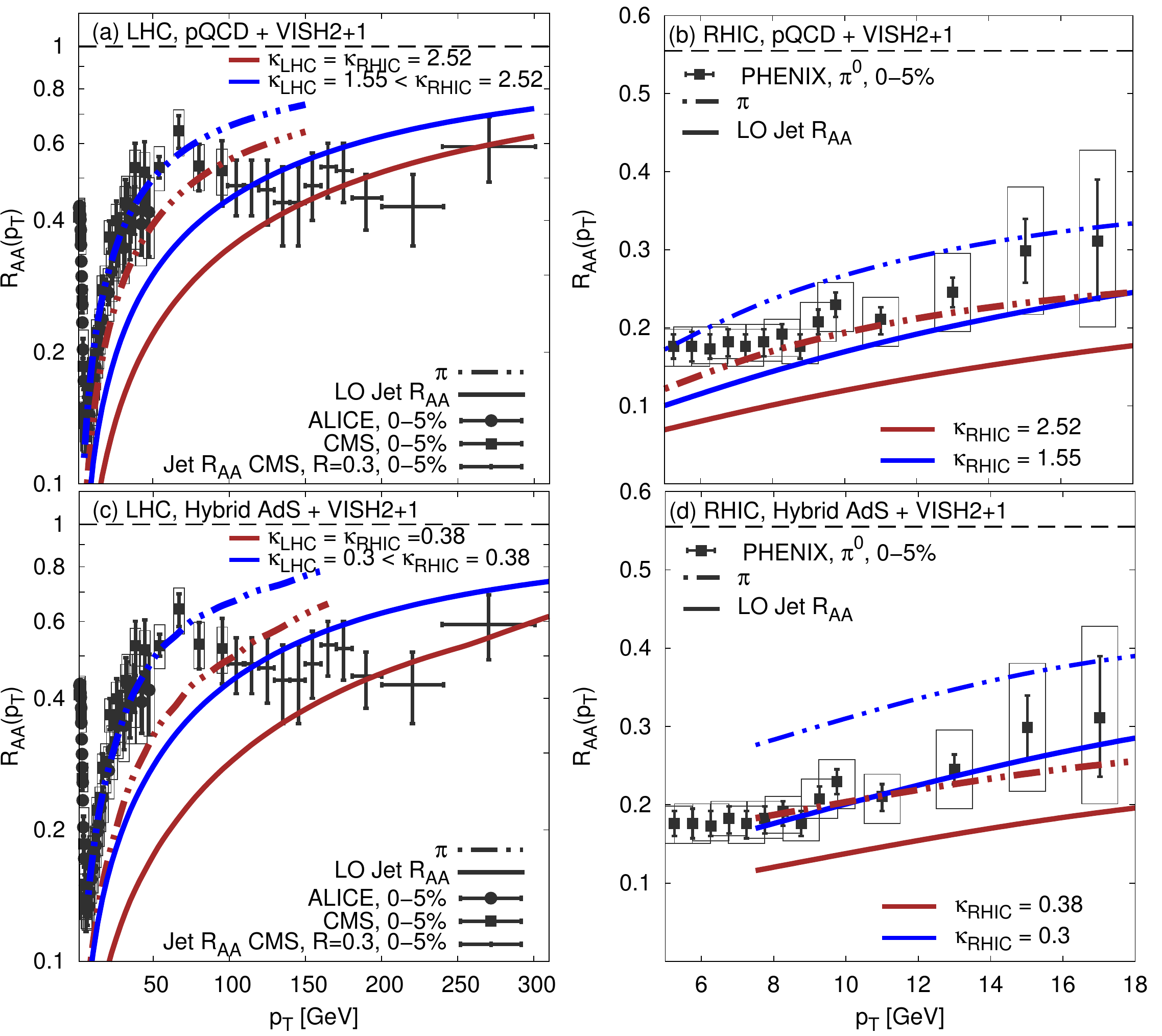}
\caption{Comparison of the pion nuclear modification factor (dashed-dotted lines) and a 
LO Jet $R_{AA}$ (solid lines) calculated via a radiative pQCD-like energy-loss 
ansatz given by $dE/d\tau=\kappa E^0 \tau^1 e^{3/4} \zeta_{0} v_f$ (upper panels)
with a constant jet-medium coupling and the hybrid strong/weak string energy-loss 
ansatz of Ref.\ \cite{Casalderrey-Solana:2014bpa} (lower panel) at LHC (left) and 
RHIC (right) energies for larger (red) and lower (blue) jet-medium couplings 
compared to the measured data \cite{data2,data3,CMS:2012rba}.
\vspace*{-0.8cm}}
\label{Fig03}
\end{center}
\end{figure*}

This jet-medium coupling was derived from the DGLV gluon number distribution 
\cite{Xu:2014tda} and is given by the analytic formul
\begin{eqnarray}
\kappa(E^2,T) &=& \alpha_S^2(E^2)\chi_T\left(f_E^2+f_E^2 f_M^2 \mu^2/E^2\right)\nonumber\\
&& - (1-\chi_T)(f_M^2 + f_E^2 f_M^2\mu^2/E^2)
\,.
\label{Eq3}
\end{eqnarray}
It includes a running coupling $\alpha_S(E^2)=1/(c+9/4\pi\,\log(E^2/T_c^2))$ with
$c=1.05$, the Polyakov-loop suppression of the color-electric scattering \cite{Hidaka:2008dr}
via $\chi_T = c_q L + c_g L^2$ with pre-factors $c_q,c_g$ for quarks and gluons, and 
the Polyakov loop $L<1$ as parametrized from lattice QCD, as well as an 
enhancement of scattering due to the magnetic monopoles near the critical temperature
$T_c$ also derived from lattice QCD \cite{Nakamura:2003pu}. This temperature and 
energy-dependent jet-medium coupling shows an effective running 
as it decreases with temperature. 

We included the above jet-medium coupling $\kappa=\kappa(E^2,T)$ in our jet-energy loss approach \cite{us}. 
The result is shown in Fig.\ \ref{Fig02}, again for the hydrodynamic background 
VISH2+1 (solid lines) and a medium determined via the parton cascade BAMPS 
(dashed lines). For comparison, we depict the results from CUJET3.0
\cite{Xu:2014tda}.

As in Fig.\ \ref{Fig01}, the ion nuclear modification factor is well 
described both at RHIC and LHC. The high-$p_T$ elliptic flow, however, increases drastically 
below $p_T \sim 20$~GeV as compared to Fig.\ \ref{Fig01}, especially for the BAMPS background which already includes 
non-equilibrium effects \cite{BAMPS}. 

Finally, we compare results from the linear pQCD approach of Eq.\ (\ref{Eq1}) 
with the highly non-linear hybrid AdS holographic model of jet-energy loss,
see Eq.\ (\ref{Eq2}). We compare the pion nuclear modification factor and
an idealized LO Jet $R_{AA}$ given by
\begin{eqnarray}
{\rm LO}\;{\rm Jet}\; R_{AA} = \frac{R_{AA}^g\,d\sigma_g(p_T)+R_{AA}^q\,d\sigma_q(p_T)}{d\sigma_g(p_T)+d\sigma_q(p_T)}\,.
\end{eqnarray}
Naturally, this LO Jet $R_{AA}$ represents a reconstructed jet with vanishing 
cone radius and is only a lower bound for the NLO Jet $R_{AA}$ with jet-cone radii $R>0$.

Fig.\ \ref{Fig03} shows this comparison at LHC (left) and RHIC (right) 
energies for two different jet-medium couplings that are treated as constants: 
A larger one (red) fitted to the pion $R_{AA}$ data (dashed-dotted lines) at RHIC 
and a lower one (blue) fitted the pion $R_{AA}$ data at LHC. 

To guide the eye, we include the reconstructed Jet $R_{AA}$ from CMS \cite{CMS:2012rba} with $R=0.3$
in Fig.\ \ref{Fig03}. The solid blue lines for the LO Jet $R_{AA}$ in the left panels of Fig.\ \ref{Fig03} 
lie in the same ballpark as the experimental data. Fragmenting this result to pions (dashed-dotted lines) 
leads to an $R_{AA}$ that reproduces the measured pion nuclear modification factor at LHC. 
A straight extrapolation of this results to RHIC energies shows that the LO Jet $R_{AA}$ for {\it the same}
jet medium couplings lie on top of the measured {\it pion} nuclear modification factor. 
However, fragmenting this result to pions leads to a $R^{\pi}_{AA}$ that is larger than 
the measured data at RHIC. 

Larger jet-medium couplings (red lines), on the other hand, describe the {\it pion} nuclear
modification factor at RHIC for the pQCD scenario and the LO Jet $R_{AA}$ at LHC is again close 
to the experimental data. The pion nuclear modification factor at LHC, however, only touches the lower bound 
of present error bars. In case of the Hybrid AdS energy loss the results always only
touch the lower end of the experimental error bars.

Fig.\ \ref{Fig03} demonstrates that the results for the pQCD and the Hybrid
AdS energy-loss including a Bragg peak are remarkably similar. Thus, unfortunately,
neither the pion nor a LO Jet $R_{AA}$ are sensitive to the difference in the path-length 
between pQCD and AdS models.

\section{Conclusions}

We compared the measured data on the nuclear modification factor for pions and reconstructed jets 
as well as on the high-$p_T$ elliptic flow at RHIC and LHC energies to results 
obtained by a linear pQCD and a highly non-linear hybrid AdS holographic model of 
jet-energy loss. We found that the simultaneous description of the $R_{AA}$
and $v_2$ requires a jet-medium coupling that depends on the energy of the
jet, the temperature of the medium \cite{us,Xu:2014tda}, and non-equilibrium effects
around the phase transition. We also contrasted a hydrodynamic
background (VISH2+1) \cite{VISH2+1} with a medium obtained from the parton cascade BAMPS
\cite{BAMPS} and showed that the influence of the underlying bulk medium considered 
is suprisingly small. Unfortunately, neiter the pion nor the LO Jet $R_{AA}$ 
are sensitive to the difference in the path-length between pQCD and AdS models.

\section{Acknowledgement}
This work was supported in part through the Bundesministerium f\"ur Bildung
und Forschung, the Helmholtz International Centre for FAIR within the framework 
of the LOEWE program (Landesoffensive zur Entwicklung 
Wissenschaftlich-\"Okonomischer Exzellenz) launched by the State of Hesse, 
the US-DOE Nuclear Science Grant No.\ DE-AC02-05CH11231 within the framework 
of the JET Topical Collaboration, and the US-DOE Nuclear Science Grant No.\
DE-FG02-93ER40764. Numerical computations have been performed at the Center for 
Scientific Computing (CSC).


\begin{thebibliography}{00}

\bibitem{data1}
A.~Adare {\it et al.} [PHENIX Collaboration], Phys.\ Rev.\ C {\bf87}, 034911 (2013);
Phys.\ Rev.\ Lett.\  {\bf 105}, 142301 (2010).

\bibitem{data2}
B.~Abelev {\it et al.}  [ALICE Collaboration], Phys.\ Lett.\ B {\bf 720}, 52 (2013);
Phys.\ Lett.\ B {\bf 719}, 18 (2013).

\bibitem{data3}
S.~Chatrchyan {\it et al.}  [CMS Collaboration], Eur.\ Phys.\ J.\ C {\bf 72}, 1945 (2012);
Phys.\ Rev.\ Lett.\  {\bf 109}, 022301 (2012).

\bibitem{data4}
G.~Aad {\it et al.}  [ATLAS Collaboration], Phys.\ Lett.\ B {\bf 707}, 330 (2012).

\bibitem{us}
B.~Betz and M.~Gyulassy, arXiv:1503.07671 [hep-ph];
JHEP {\bf 1408}, 090 (2014) [Erratum-ibid.\  {\bf 1410}, 043 (2014)];
Phys.\ Rev.\ C {\bf 86}, 024903 (2012).

\bibitem{VISH2+1}
C.~Shen, U.~Heinz, P.~Huovinen and H.~Song, Phys.\ Rev.\ C {\bf 84}, 044903 (2011);
 Phys.\ Rev.\ C {\bf 82}, 054904 (2010).

\bibitem{BAMPS}
J.~Uphoff {\it et al.}, Phys.\ Rev.\ Lett.\  {\bf 114}, no. 11, 112301 (2015);
Z.~Xu and C.~Greiner, Phys.\ Rev.\ C {\bf 71}, 064901 (2005).

\bibitem{Casalderrey-Solana:2014bpa} 
J.~Casalderrey-Solana {\it et al.}, JHEP {\bf 1410}, 19 (2014).

\bibitem{Chesler:2014jva} 
P.~M.~Chesler and K.~Rajagopal, Phys.\ Rev.\ D {\bf 90}, 025033 (2014).

\bibitem{Betz:2008ka} 
B.~Betz {\it et al.}, Phys.\ Rev.\ C {\bf 79}, 034902 (2009).

\bibitem{Ficnar}
A.~Ficnar, S.~S.~Gubser and M.~Gyulassy, arXiv:1404.0935 [hep-ph];
Phys.\ Lett.\ B {\bf 738}, 464 (2014);
A.~Ficnar and S.~S.~Gubser, Phys.\ Rev.\ D {\bf 89}, 026002 (2014).

\bibitem{Xu}
J.~Xu, A.~Buzzatti and M.~Gyulassy, JHEP {\bf 1408}, 063 (2014).

\bibitem{Molnar}
D.~Molnar and D.~Sun, arXiv:1305.1046 [nucl-th]; Nucl.\ Phys.\ A {\bf 910-911}, 486 (2013).

\bibitem{Xu:2014tda} 
J.~Xu, J.~Liao and M.~Gyulassy, Chin.\ Phys.\ Lett.\  {\bf 32}, no. 9, 092501 (2015).

\bibitem{Hidaka:2008dr} 
Y.~Hidaka and R.~D.~Pisarski, Phys.\ Rev.\ D {\bf 78}, 071501 (2008).

\bibitem{Nakamura:2003pu} 
A.~Nakamura, T.~Saito and S.~Sakai, Phys.\ Rev.\ D {\bf 69}, 014506 (2004).

\bibitem{CMS:2012rba} 
CMS Collaboration [CMS Collaboration], CMS-PAS-HIN-12-004.

\end{thebibliography}
\end{document}